\documentclass[pra,aps,showpacs,superscriptaddress]{revtex4}

\newcommand{\half}{\frac{1}{\sqrt{2}}}
\newcommand{\halv}{\frac{1}{2}}
\begin{document}
\title{Local  quantum mechanics  and  superquantum correlations  imply
  superluminal         signaling}         

\author{S.         Aravinda}

\affiliation{Poornaprajna    Institute    of   Scientific    Research,
  Sadashivnagar, Bangalore, India.}

\author{R.     Srikanth}\email{srik@poornaprajna.org}  

\affiliation{Poornaprajna    Institute    of   Scientific    Research,
  Sadashivnagar, Bangalore, India.}

\affiliation{Raman   Research  Institute,   Sadashivnagar,  Bangalore,
  India.}

\begin{abstract}
It  has been shown  that local  quantum measurements  and no-signaling
imply quantum  correlations (Barnum et  al.  Phys.  Rev.   Lett., 104,
140401 (2010)).   This entails that superquantum  correlations will be
superluminaly  signaling if  we admit  the validity  of  local quantum
mechanics.  We present a  specific, simple instance of this situation.
Our  result  also has  the  interpretation  that  replacing the  local
(trivial) dynamics in a generalized non-signaling theory (GNST) by one
that  admits local  unitaries, results  in a  probability distribution
that  may  be   positive  and  normalization-preserving  but  violates
no-signaling, and is thus inconsistent within the formalism.
\end{abstract}

\maketitle

\section{Introduction}

Quantum correlations  are nonlocal in that they  can violate Bell-type
inequalities \cite{bell,chsh} that  classical correlations cannot, but
are themselves bounded by the Tsirelson bound \cite{cir}. It was shown
in Ref.   \cite{prb} that there exist  super-quantum correlations that
can  exceed  this  bound,  violating  Bell-type  inequalities  to  the
algebracally   maximally  allowed   degree,   while  still   remaining
non-signaling. This meant  that nonlocality and relativistic causality
were insufficient as axioms  to derive quantum mechanics, and provoked
the   natural  question   of  why   such   super-quantum  correlations
(Popescu-Rohrlich (PR) boxes and ``nonlocal boxes'') were not observed
in  Nature. This  has  spurred a  great  deal of  research devoted  to
elucidating   the  nature   of  non-signaling   nonlocal  correlations
\cite{bar,piro,gis,rel,qua,gar,resource}     in     general,     their
applications  to cryptography  \cite{gis}, to  simulating correlations
\cite{anne},  and  an  examination  of  possible  or  actual  physical
principles that may preclude the  occurance of nonlocal boxes, such as
the non-triviality of communication complexity \cite{van}, information
causality \cite{inf}, or bounds  implied by the Heisenberg uncertainty
\cite{weh}.  Ref.  \cite{indra}  shows  that  PR  boxes  can  lead  to
superluminal  signaling when one  of the  input bits  has access  to a
closed time curve (CTC).

A PR-box is described by the action
\begin{equation}
P(a,b|A,B) = \left\{ 
\begin{array}{l l}
  \frac{1}{2} & \quad \mbox{if $ a \oplus b =A \cdot  B$ }\\
  0 & \quad \mbox{otherwise}\\ \end{array} \right.
\label{eq:nlb}
\end{equation}
where $A$ and  $B$ are the respective (binary)  inputs of two players,
Alice and  Bob, and $a$ and  $b$ their respective  outputs. By design,
the  box (\ref{eq:nlb}),  or any  other equivalent  obtained  by local
relabelling, satisfies the no-signaling condition
\begin{equation}
\sum_b P(a,b|A,B=0) = \sum_b P(a,b|A,B=1) \equiv P(a|A).
\label{eq:nosig}
\end{equation}

In order to  be able to answer the question raised  above, it would be
useful  to set  quantum theory  within the  framework of  more general
probability  theories,  a problem  considered  by several  researchers
recently  \cite{bar,gis,rel,weh}. In  Ref.  \cite{qua},  it  was shown
that  assuming that quantum  mechanics holds  good locally  -- meaning
that  probabilities for  any  local measurement  from reduced  density
operators are obtained in the standard non-contextual way --, and that
the  no-signaling  principle  is   true,  then  the  correlations  are
necessarily  quantum mechanical.  This  entails that  if super-quantum
correlations  exist,  then  either  local quantum  mechanics  will  be
invalidated   or  the   no-signaling  principle   will   be  violated.
Therefore, if  quantum mechanics is valid  locally, then super-quantum
correlations necessarily leads  to superluminal signaling.  We present
a   specific   instance   of   this   situation   in   the   following
Section.   Assuming   local   validity   of  quantum   mechanics   and
no-signaling,  this  provides another  perspective  into why  nonlocal
boxes are not found in Nature.
 
\section{Superpositional inputs of local quantum mechanics}

Traditionally,  the  inputs  $A$  and  $B$  to  the  PR-box  represent
measurement choices,  but here  it will be  helpful to regard  them as
abstract  (binary)   variables  \cite{piro}.   In   that  spirit,  and
considering  that  an experiment  here  is  performed  in a  (quantum)
physical world, if  Alice inputs 0 and Bob inputs  1, then their joint
input can surely be represented as the state vector $|01\rangle \equiv
|0\rangle|1\rangle  \equiv  |\Phi_0\rangle$   living  in  an  abstract
Hilbert space ${\cal H}_A \otimes {\cal H}_B$, and represented here in
the computational  basis. That is, the  inputs can be  considered as a
two-qubit state rather than  two qubit-measurements.  The nonlocal box
(\ref{eq:nlb}) acting on $|\Phi_0\rangle$ yields
\begin{equation}
|\Psi_0) = \halv(|00\rangle \odot |11\rangle),
\label{eq:psi0}
\end{equation}
where $\odot$ represents an incoherent sum, and Eq. (\ref{eq:psi0}) is
a vector-like representation for the resulting density operator.

If  Alice rotates  her input  state  even slightly  using the  unitary
$\left(  \begin{array}{cc} \cos\theta  & \sin\theta  \\  -\sin\theta &
  \cos\theta \end{array}\right)$, then the joint input to the nonlocal
box is the  superposition $|\Phi_1\rangle \equiv \half(\cos|0\rangle +
\sin|1\rangle)|1\rangle$.   We  consider a  PR-box  that accepts  such
superposition (to whatever small degree) states as input data and acts
linearly on them.

The  application  of Eq.   (\ref{eq:nlb})  to vector  $|\Phi_1\rangle$
input to such a PR-box produces the mixture
\begin{equation}
  |\Psi_1) \equiv \frac{1}{2}\left(\cos\theta(|00\rangle  \odot |11\rangle) 
   + 
  \sin\theta(|01\rangle \odot |10\rangle)\right).
\label{eq:strang}
\end{equation}  
It may  be noted that $|\Psi_1)$ is  a coherent \textit{superposition}
of a  \textit{classical} mixture.  Although  perhaps somewhat unusual,
such superpositions occur in standard quantum mechanics.  For example,
consider  an atom with  both its  ground state  and the  first excited
state having two-fold degeneracy.  A  basis for the ground subspace is
denoted by $\{|g,0\rangle, |g,1\rangle\}$,  and similarly that for the
excited  subspace  by  $\{|e,0\rangle, |e,1\rangle\}$.   Two  unbiased
random classical  bits, $G$  and $E$, are  generated and input  into a
state   preparation  apparatus,   which   outputs  the   superposition
$\half(|g,G\rangle + |e,E\rangle)$.  The output state may equivalently
be  written  as  $\half\left(\halv(|g,0\rangle  \odot  |g,1\rangle)  +
\halv(|e,0\rangle    \odot    |e,1\rangle)\right)$,    similarly    to
Eq. (\ref{eq:strang}).

The superposition operator `+' can be interpreted as a logical AND and
operator `$\odot$' as logical OR, allowing us to distribute the latter
through the former, and thus re-write Eq. (\ref{eq:strang}) as
\begin{eqnarray}  
|\Psi_1)    &=&    \frac{1}{4}    \left(    (\cos\theta|00\rangle    +
\sin\theta|01\rangle)        \odot       (\cos\theta|00\rangle       +
\sin\theta|10\rangle)        \odot       (\cos\theta|11\rangle       +
\sin\theta|01\rangle)        \odot       (\cos\theta|11\rangle       +
\sin\theta|10\rangle)       \right)       \nonumber       \\       &=&
\frac{1}{4}\left(|0\rangle  |\phi_0  \rangle  \odot |\phi_0\rangle  |0
\rangle    \odot   |\phi_1\rangle    |1   \rangle    \odot   |1\rangle
|\phi_1\rangle\right)
\end{eqnarray}
The    second   register    is   in    the   state    $\rho_1   \equiv
\frac{1}{2}\left(    \begin{array}{cc}   1    &   \cos\theta\sin\theta
  \\ \cos\theta\sin\theta & 1 \end{array}\right)$.  On the other hand,
this register  in Eq. (\ref{eq:psi0})  is in the state  $\rho_0 \equiv
\frac{1}{2}\left(    \begin{array}{cc}    1     &    0    \\    0    &
  1  \end{array}\right)$.   Thus  by   choosing  her  input  to  be  a
non-application  or  application   of  the  superposition,  Alice  can
transmit a probabilistic superluminal bit to Bob. By sufficiently many
repetitions,  the  transmission probability  can  be made  arbitrarily
close to 1, even for arbitrarily small superposition ($\theta \ll 1$).

\section{Poverty of the dynamics of the nonlocal box world}

Our result  can also  be interpreted in  another way.  Although  it is
known  that  the  set  of  allowed  dynamical  transformation  in  the
box-world theories,  and more  generally, GNST \cite{bar},  is trivial
\cite{gross} (convex  combinations of  outcome relabelling), it  is an
interesting  question in  what way  a given  disallowed transformation
leads  to invalid  states.   For  example, it  could  lead to  invalid
results by virtue of  negative probabilities or loss of normalization.
Our result shows that  even slight superpositions lead to superluminal
signaling, though  they remain positive  and normalization preserving.
This more  general than trivial  dynamics is thus  inconsistent within
the no-signaling framework.   While this signaling property presumably
makes it physically  unviable, it may nevertheless be  relevant to the
question of deriving  quantum mechanics from computational assumptions
\cite{sri}.

\section{Conclusions and Discussions}

We provide a  concrete instance of local quantum  mechanics taken with
superquantum  correlations leading to  superluminal signaling.   It is
worth  noting that  the two  assumptions that  rule  out super-quantum
correlations, namely  that of local validity of  quantum mechanics and
of no-signaling, are themselves  closely related in quantum mechanics.
In   Ref.     \cite{himanshu},   it   was    shown   that   Gleasonian
non-contextuality, a  facet of local quantum  mechanics, together with
the assumption of tensor product structure, implies no-signaling. This
provides yet another insight into  why nonlocal boxes are not found in
Nature.

\acknowledgments

SA  acknowledges  support  through   the  INSPIRE  fellowship  by  the
Department of Science and Technology, Govt. of India.

\end{document}